\documentstyle[preprint,aps]{revtex}

\begin{document}
\bibliographystyle{prsty}
\draft

\title{Grover search with pairs of  trapped ions}
\author{Mang Feng
\thanks{Electronic address: feng@mpipks-dresden.mpg.de}} 

\address{Max-Planck Institute for the Physics of Complex Systems,\\
N$\ddot{o}$thnitzer Street 38, D-01187 Dresden, Germany}

\date{\today}

\maketitle

\begin{abstract}

The desired interference required for quantum computing may be modified by
the wave function oscillations for the implementation 
of quantum algorithms[Phys.Rev.Lett.84(2000)1615]. To diminish such detrimental
 effect, we propose a scheme with trapped ion-pairs being qubits and apply the 
 scheme to the Grover search. It can be found that our scheme can not only 
carry out a full Grover search, but also meet the requirement for the scalable 
hot-ion quantum computing. Moreover, the ion-pair qubits in our scheme are more 
robust against the decoherence and the dissipation caused by the environment than
single-particle qubits proposed before.
\end{abstract}
\vskip 1cm

\narrowtext

\section{Introduction}

Since Shor's discovery$^{[1]}$ of the quantum algorithm for factoring large 
integers, much progress has been made in the field of the quantum computing. It 
has been 
generally considered that the quantum computer distinguishes the classical 
computer in the capabilities
to operate quantum mechanically on superpositions of quantum states and
to exploit resulting interference effects. With these capabilities,
quantum computers can outperform classical ones in solving classically
intractable problem$^{[1,2]}$ or solving tractable problems more 
rapidly$^{[3]}$. It has been proposed that several physical systems can be 
used for the quantum computing, such as the ion trap, nuclear magnetic 
resonance(NMR) system, atom-cavity interaction and so on. However,
up to now, most experimental demonstrations concerning the quantum algorithm and
quantum communication have been solely performed with NMR technique$^{[4,5]}$
due to its technique maturity. Meanwhile there are intensive dispute$^{[6]}$ 
for whether the genuine quantum computing was made in the NMR system since the 
manipulation with NMR technique is applied on the bulk molecules instead of the individual
molecule. In contrast, in the ion trap quantum computing, the 
manipulation is indeed performed on individual trapped ion by quantum level,
and the coupling between the electronic states of the ion and its vibrational 
motion is made by the laser fields. Since the success of first 
experiment$^{[7]}$ of
two-qubit controlled-NOT with a single ultracold $Be^{+}$ based on the
proposal by Cirac and Zoller$^{[8]}$, many related theoretical schemes$^{[9]}$
have been put forward, and the experimental progress$^{[10,11]}$ in this 
respect has also been made.
However, it is hard to achieve the entanglement of large numbers of trapped
ions because the experiment relied on the particular behavior of the 
ions$^{[10]}$.
Recently, an approach with bichromatic field$^{[12]}$ was proposed, which leads
to the success of entanglement of four trapped ions$^{[13]}$. 
In that proposal, two identical two-level ions in the string are both illuminated with 
two lasers of different frequencies $\omega_{1,2}=\omega_{eg}\pm\delta$, where 
$\omega_{eg}$ is the resonant transition frequency of the ions, and $\delta$ 
the detuning, not far from the trap frequency $\nu$. With the choice of laser 
detunings the only energy conserving transitions are from $|ggn>$ to $|een>$ 
or from $|gen>$ to $|egn>$, where the first(second) letter denotes the 
internal state $e$ or $g$ of the $i^{th}$(j$^{th}$) ion and $n$ is the quantum 
state for the vibrational state of the ion. That is to say, the
states $|ggn>$ and $|een>$, separated by $\omega_{1}+\omega_{2}$ are
resonantly coupled and so are the degenerated states $|egn>$ and $|gen>$.
 As we consider $\nu-\delta\gg\eta\Omega$ with $\eta$ being
the Lamb-Dicke parameter and $\Omega$ the Rabi frequency, there is only 
negligible population being transferred to the intermediate states with 
vibrational quantum number $n\pm 1$. It has been proven that this 
two-photon process is nothing to do with the vibrational state
$|n>$. So the quantum computing with such configuration is valid even for
the hot ions.

As we know, the implementation of the quantum computing is based on two basic 
operations$^{[14]}$. One is the single-qubit rotation, and the other 
the two-qubit operation. The suitable composition of such two operations 
will in principle carry out any quantum computing operation we wanted. However, 
the quantum computing is implemented on the superposition of
eigenstates of the Hamiltonian. According to the Schr\"odinger equation,
during the time interval $t$, each quantum state $\Psi_{i}$ acquires a phase
$-E_{i}t$, where $E_{i}$ is the eigenenergy of the state 
$\Psi_{i}$(supposing $\hbar=1$). Thus any delay time between the operations 
will produce unwanted different phases in different quantum
states$^{[15,16]}$, which modifies the quantum
interference of an ideal quantum computing, and spoils the correct results we
desired. How to avoid this detrimental effect? 
The authors of Ref.[15] proposed an ideal solid-state qubits model making use 
of controllable low-capacitance Josephson junction to avoid this undesired
phase evolution, in which energy splitting  between logic
states can be tuned to be zero during the delay time. Ref.[16] carried out a
general consideration on this problem and suggested to
 use stably continuous reference oscillations with
the resonant frequency for each quantum transition in the process of the 
quantum computing. We consider that the most efficient approach is to use the
degenerated states to be the logic states, which can transfer the relative phases to
a global one and the errors caused by the relative phases would be
eliminated completely.

In this contribution, we will demonstrate a scheme to pair the trapped 
ions to be a qubit for eliminating the 
detrimental effect referred to above. Our proposal is based on the hot-ion 
quantum computing model of Ref.[12], by choosing the transition paths from 
$|egn>$ to $|gen>$, and setting $|eg>=|0>$ and $|ge>=|1>$. As the
qubits $|0>$ and $|1>$ are degenerated in energy, no unwanted relative
phases will appear  in the delay
time between any two of the operations. We will first carry out a two-qubit 
Grover search with our scheme, and then extend the scheme to the more-qubit cases.
Finally, a discussion will be made for the implementation of a full Grover search as well as
the advantage and limitation of our work.

\section{Two qubits Grover search with trapped ion-pairs}

In the Lamb-Dicke limit($\eta\ll 1$) and weak excitation
regime($\Omega<\nu$), we may obtain the time evolution of the states from the 
second order perturbation theory with the definition of effective Rabi 
frequency $\tilde{\Omega}=-\frac {(\Omega\eta)^{2}}{2(\nu-\delta)}$$^{[12]}$,
$$\hat{U}|1>=\cos(\frac {\tilde{\Omega}T}{2})|1>-i\sin(\frac {\tilde{\Omega}T}
{2})|0>,$$
\begin{equation}
\hat{U}|0>=\cos(\frac {\tilde{\Omega}T}{2})|0>-i\sin(\frac {\tilde{\Omega}T}{2})|1>.
\end{equation}
Setting $|1>=\pmatrix {0\cr 1}$ and $|0>=\pmatrix {1\cr 0}$, we obtain 
$\hat{U}=\hat{U}(\theta)$
$=\pmatrix{\cos\theta&-i\sin\theta \cr -i\sin\theta&\cos\theta}$ with 
$\theta=\frac {\tilde{\Omega}T}{2}$. To construct a quantum computing
model, what we need to do in the following is to find a suitable two-qubit
operation, like controlled-NOT gate, i.e., if and only if the first ion and 
the second ion are respectively in states $|g>$ and $|e>$, the third and 
fourth ions will be flipped, making
$|eg>\rightarrow |ge>$ and $|ge>\rightarrow |eg>$. 

Please note that $\hat{U}(\theta)$ is not the general form of  Walsh-Hadamard 
gate although it
plays a similar role to the Walsh-Hadamard gate. So we introduce a two-qubit
operation $\hat{M}^{(2)}_{1}=\pmatrix {1&0&0&0\cr 0&1&0&0\cr 0&0&0&-i\cr 0&0&i&0}$,
which plays similar role to the controlled-NOT gate. In what follows, we implement 
a two-qubit Grover search with above operations as an example. The most 
efficient Grover search includes three kinds of operations in an
iteration(i.e., a searching step)$^{[3]}$, (i) preparing a
superposition of states with equal amplitude; (ii) inverting the amplitude 
of the marked state; (iii) performing a diffusion transform $\hat{D}$,
i.e., the inversion about average(IAA) operation, with $\hat{D}_{ij}=\frac {2}{N}$
for $i\neq j$ and $N=2^{q}$(q being the number of the qubits), and 
$\hat{D}_{ii}=-1+\frac {2}{N}$. With our method, we first
prepare two ion-pairs to the states $|ge>_{1}|ge>_{2}$, i.e.,
$|1>_{1}|1>_{2}$(labeled as $|11>$ for simplicity in the following), without 
consideration of the vibrational states.
Then $\hat{U}(\frac {7\pi}{4})$ will be performed on the two pairs
simultaneously, we obtain
\begin{equation}
|\Psi_{1}>=\hat{W}_{2}\pmatrix{0\cr 0\cr 0\cr 1}=\frac {1}{2}
\pmatrix{-1\cr i\cr i\cr 1}
\end{equation}
with
 $$\hat{W}_{2}=\frac {1}{\sqrt{2}}\pmatrix{1&i\cr i&1}\otimes
\frac {1}{\sqrt{2}}\pmatrix{1&i\cr i&1}= 
 \frac {1}{2}\pmatrix{1&i&i&-1\cr i&1&-1&i\cr i&-1&1&i\cr -1&i&i&1}.
$$
As $\hat{U}(\theta)$ is not the Walsh-Hadamard gate, $|\Psi_{1}>$ is not the
superposition of states as the original Grover method required$^{[3]}$. 
But that does not matter. We can still continue the procedure in the Grover 
search. Supposing that the marked state is $|11>$, we have to invert the
amplitude of this state, that is,
\begin{equation}
|\Psi_{2}>=\hat{P}^{(2)}_{1}\frac {1}{2}\pmatrix{-1\cr i\cr i\cr 1}
=\frac {1}{2}\pmatrix{-1\cr i\cr i\cr -1}
\end{equation}
where $\hat{P}^{(2)}_{1}=\hat{V}_{2}^{-1}\hat{M}^{(2)}_{1}\hat{V}_{2}=
\pmatrix {1&0&0&0\cr 0&1&0&0\cr 0&0&1&0\cr 0&0&0&-1}$ with 
$\hat{V}_{2}=\pmatrix{1&0\cr 0&1}$$\otimes
\frac {1}{\sqrt{2}}\pmatrix{1&i\cr i&1}$. The operation $\hat{V}_{2}$ means
that $\hat{U}(\frac {7\pi}{4})$ is performed on the second pair,
whereas no operation on the first pair.
Finally, the IAA operation in the Grover search
can be realized by the operation
\begin{equation}
\hat{D}_{2}=\hat{W}_{2}\hat{P}^{(2)}_{1}\hat{W}_{2}=\frac {1}{2}
\pmatrix{-1&i&i&-1\cr i&1&-1&-i\cr i&-1&1&-i\cr 
-1&-i&-i&-1}.
\end{equation}
It is easily found that 
\begin{equation}
|\Psi_{3}>=\hat{D}_{2}|\Psi_{2}>=\pmatrix{0\cr 0\cr0\cr1},
\end{equation}
which means that the state $|11>$ has been found out. For further search, we
may find that the state $(0001)^{-1}$ recurs every third search steps, same as
the demonstration in Ref.[4].

According to the Grover search$^{[3]}$, one can find out a certain state by
the operation of IAA as long as the amplitude of that state has been inverted. It can be found that our
scheme also meets this requirement. Defining other three two-qubit operations
to be
$\hat{M}^{(2)}_{2}=\pmatrix {1&0&0&0\cr 0&1&0&0\cr 0&0&0&i\cr 0&0&-i&0}$,
$\hat{M}^{(2)}_{3}=\pmatrix {0&-i&0&0\cr i&0&0&0\cr 0&0&1&0\cr 0&0&0&1}$,
and $\hat{M}^{(2)}_{4}=\pmatrix {0&i&0&0\cr -i&0&0&0\cr 0&0&1&0\cr 0&0&0&1}$, the
inversion operations will be 
$\hat{P}^{(2)}_{2}=\hat{V}_{2}^{-1}\hat{M}^{(2)}_{2}\hat{V}_{2}=
\pmatrix {1&0&0&0\cr 0&1&0&0\cr 0&0&-1&0\cr 0&0&0&1}$,
$\hat{P}^{(2)}_{3}=\hat{V}_{2}^{-1}\hat{M}^{(2)}_{3}\hat{V}_{2}=\pmatrix {1&0&0&0\cr 
0&-1&0&0\cr 0&0&1&0\cr 0&0&0&1}$ and $\hat{P}^{(2)}_{4}=\hat{V}_{2}^{-1}\hat{M}^{(2)}_{4}
\hat{V}_{2}=\pmatrix {-1&0&0&0\cr 0&1&0&0\cr 0&0&1&0\cr 0&0&0&1}$.
If we want to find out a certain $ith$ state, the search process will be the
same as the above, except that a specific $\hat{P}^{(2)}_{i}$ operation is made to
invert the amplitude of the $ith$ state. That is to say, no matter which 
state is to be searched, the IAA operation is still 
$\hat{D}_{2}=\hat{W}_{2}\hat{P}^{(2)}_{1}\hat{W}_{2}$. But we need a specific 
operation $\hat{P}^{(2)}_{i}$ to invert the amplitude of the $ith$ state before each 
IAA operation.
 
\section{More qubits Grover search}

Along the idea in last section, we can extend the technique to the many-qubit cases for 
the Grover search. 
For a q-qubit case, we should first construct the Walsh-Hadamard gate as follows
\begin{equation}
\hat{W}_{q}=\frac {1}{\sqrt{2}}\pmatrix{1&i\cr i&1}\otimes\cdots\otimes
\frac {1}{\sqrt{2}}\pmatrix{1&i\cr i&1}
\end{equation}
where $\cdots$ represents the tensor product of q-2 terms, and then the IAA 
operation is 
$\hat{D}_{q}=\hat{W}_{q}\hat{P}^{(q)}_{1}\hat{W}_{q}$ with the $2^{q}\times 2^{q}$ matrix 
$\hat{P}^{(q)}_{1}=
\pmatrix {1&0&\ldots &0&0\cr  
          \vdots &\vdots &\vdots &\vdots &\vdots\cr
          0&0&\ldots &1&0\cr
          0&0&\ldots&0&-1}$. $\hat{P}^{(q)}_{1}$ can be
implemented by the operations $\hat{V}_{q}^{-1}\hat{M}^{(q)}_{1}\hat{V}_{q}$ with $2^{q}\times 2^{q}$ matrix 
$\hat{M}^{(q)}_{1}=\pmatrix {1&0&\ldots &0&0&0\cr  
          \vdots &\vdots &\vdots &\vdots &\vdots\cr
          0&0&\ldots &1&0&0\cr
          0&0&\ldots &0&0&-i\cr
          0&0&\ldots&0&i&0}$  and $\hat{V}_{q}=\underbrace{\pmatrix{1&0\cr 0&1}\otimes\cdots\otimes
\pmatrix{1&0\cr 0&1}}_{q-1~ terms}\frac {1}{\sqrt{2}}\pmatrix{1&i\cr i&1}$.
To invert the amplitude of the marked state, we need not only $\hat{P}_{1}^{(q)}$, but also
$\hat{P}^{(q)}_{i}(i=2,3,\cdots,2^{q})$. These matrixes can be constructed similarly. With the 
operations $\hat{W}_{q}$, $\hat{P}^{(q)}_{i}$ and $\hat{D}_{q}$, we can carry out  the many-qubit Grover 
search.

However we have no way to present a general expression for the result of each iteration 
of a many-qubit Grover search with the present scheme as did in Ref.[17]. So we only 
investigated specifically the 2$\sim$5 qubits Grover search with the scheme by means of 
specific calculations with Mathematica. For the case of $N=3$, the Walsh-Hadamard gate
is $\hat{W}_{3}=\frac {1}{\sqrt{2}}\pmatrix{1&i\cr i&1}\otimes
\frac {1}{\sqrt{2}}\pmatrix{1&i\cr i&1}\otimes\frac {1}{\sqrt{2}}\pmatrix{1&i\cr i&1}$
and the IAA operation is 
$$D_{3}=\frac {1}{8}\pmatrix{1&-i&-i&-1&-i&-1&-1&-3i\cr
                         -i&-1&-1&i&-1&i&-3i&1 \cr
                         -i&-1&-1&i&-1&-3i&i&1 \cr
                         -1&i&i&1&-3i&1&1&-i \cr
                         -i&-1&-1&-3i&-1&i&i&1 \cr
                         -1&i&-3i&1&i&1&1&-i \cr
                         -1&-3i&i&1&i&1&1&-i \cr
                         -3i&1&1&-i&1&-i&-i&-1 \cr}.$$

We use $|111>$ as the initial state, and suppose the marked state
is also $|111>$. The result of each iteration of the search is
plotted in Fig.1. We may find that, compare to the standard iteration process
of the Grover search$^{[3,17]}$, although we also have large probabilities (i.e. $\geq
50\%$) to find out the desired state successfully within $\sqrt{8}$ steps, 
the searching result with our scheme is no longer strictly periodic. The numerical result 
also showed that, the amplitude of the state $|000>$ always equals that of $i|111>$
in the searching process. It means that we will acquire two results
simultaneously by
direct measurement, and one of them has to be deleted as an undesired
solution. It adds the searching steps and will much lower the efficiency of 
the Grover search if the multiply maximal probabilities will appear in the many-qubit
Grover search. Fortunately, such problem was not found in investigating the 
4-qubit and 5-qubit cases. From Figs.2 and 3, we also know that, like
3-qubit case, the searches succeed within $\sqrt{N}$$(N=2^{4}$ and $2^{5})$ 
steps, and the results of the searches are also not strictly periodic. 
It may be speculated that the problem of multiply
maximal probabilities will no longer take place for the more qubits Grover search.
Therefore, with our scheme, the maximal number of iteration for a successful 
search is still $\sqrt{N}$ except 
the 3-qubit case which needs additonal steps for judging the correct solution.

\section{Discussion and conclusion}

Before discussing the advantages 
and limitation of our scheme, we should mention that the search we described 
above is incomplete. For a full Grover search, we must first test which states will be
the marked states in order to adjust the phases of them$^{[17-20]}$, which is also called
the process of search criterion calculation$^{[19]}$. This calculation can be implemented by 
introducing the quantum random number generator(QRNG) and some extra qubits, as well as the corresponding 
software - quantum program$^{[20]}$. The QRNG
can generate a random number in the 
binary representation. By comparing the number with the binary function needed to be satisfied,
the signs of the amplitudes of states would be determined without any external influence. It is obvious 
that the Walsh-Hadamard-type gate can be used as the QRNG, and the search criterion 
calculation can be carried out polynomially under the mechanism of quantum parallelism.  
Moreover, in a tighter analysis of the Grover search$^{[17]}$, the iteration of the search was described
strictly mathematically. It is shown that, as long as we know the total number of items in
the searched database and the number of the solutions, 
the optimal searching steps can be calculated beforehand and the search can thereby be made efficiently$^{[17,18]}$.
However, the calculation for estimating the number of the solutions is not always necessary before the search is 
implemented. In practice, we can use the straightforward algorithm proposed in the Lemma 2
of Ref.[17] by introducing some random processes, which has been proven to be polynomial. So from above discussion, 
we can know that our scheme is still pratical for the full Grover search. Actually, 
as only few-qubit quantum computing has been carried out
experimentally so far, no considerations for the search criterion calculation and multi-solution problem have been taken 
in the experimantal implementation of the search$^{[4,5]}$. The quantum computing hardware is still in its infancy.
We also note that the usefulness of the Grover search
in the practical application is  questioned$^{[19]}$. As more specific discussion along this
direction is beyond the scope of the paper, in what follows, our discussion will be restricted in analyzing
the limitation and advantage of the scheme described in Sec.II and III.

The power of the Grover search increases with the increase of the number of qubits. So
a practical scheme for the search should work in the case of large numbers of qubits.
 While with our method, we did not demonstrate a general expression for  each iteration of 
the Grover search with arbitrary numbers of qubits.  The specific numerical 
calculation  was only made for the few-qubit cases. 
Moreover, we have not found how to implement $\hat{M}^{(q)}_{i}$ $(q\geq 3)$ with 
two-qubit operations $\hat{U}$ and $\hat{M}^{(2)}_{i}$
in our scheme so far. Nevertheless, according to the general discussion in Ref.[14], we can definitely achieve
the quantum computing when we have a single-qubit and a two-qubit unitary operations. Particularly, 
from the thereom 3 in Ref.[18], we know that, as long as we have
a Walsh-Hadamard-type operation, the Grover's search can be definitely achieved. So 
we consider that our scheme can have the quadratic speedup not only for
the few-qubit cases shown specifically in last section, but also for cases with 
arbitrarily large numbers of qubits. Furthermore, we noted that, the general form of the Walsh-Hadamard gate should have the 
form of SU(2). At least the gate should be with the form of 'y-axis rotation matrix' $\pmatrix{\cos(\theta/2) & \sin(\theta/2)
\cr -\sin(\theta/2) & \cos(\theta/2)}$. However, $\hat{U}(\theta)$ in our paper is 
a 'x-axis rotation matrix'$^{[14,21]}$, a special unitary operation, we consider that it is the reason results in the 
non-periodic iteration in our scheme.  

More importantly, we can find following advantages of our scheme:

(i) the undesired phase factors produced during the delay period between any two 
operations will turn to the global phase due to the degeneracy of the logic
states, which makes the actual 
implementation of an ideal quantum computing available; 

(ii) our scheme based on Ref.[12] still 
meets the requirement of hot-ion quantum computing and scalability. As the 
vibrational states of the ions are decoupled from the internal states 
of the ions, the quantum information may be processed and transferred nearly 
safely in the subspace spanned by the internal states of the ion-pairs; 

(iii) even if we assume the decoherence will probably take place in  
the actual
ion trap experiments due to some unpredictable factors  such as the intensity
fluctuations in the Raman laser beams etc$^{[13]}$, the qubits with the ion pairs 
may be 
immune against any possible decoherence caused by the surrounding environment. 
The two ions in a pair can be assumed to be decoherenced collectively
because their distance is much smaller than the effective wave length of the 
thermal noise field$^{[22]}$. As proposed in Ref.[23], by suitably choosing 
the intensity and the phase of a driving field, such a qubit can be in a
coherent-preserving state which undergoes no decoherence even if it is
interacted with the environment. It is worth being noticed that, the recent 
experiment$^{[24]}$ with polarization
entangled states of photons has produced the coherent-preserving states. So we can
expect that such a robust state will soon be produced in the ion trap experiment;  

(vi) besides the decoherence effect, there is another detrimental effect, 
i.e., the dissipation$^{[25]}$ in the interaction between the trapped ions 
and the environment. In our scheme, the dissipation effect 
will be strongly suppressed due to the degeneracy of the logic states. 

Now we make some discussions of the more technical aspects for the
physical realization of  our scheme in the ion trap. As reported in
Ref.[13], four ultracold ions have been entangled in a linear ion trap by
using the approach of bichromatic fields, and much larger numbers of ions
can be entangled in principle with the same technique. With our scheme, we 
set the $2N$ trapped ultracold ions to be $N$ qubits, and choose $|ge>$
to be the initial state in each pair. For each ion, a very weak laser beam 
is needed to detect the quantum jumps in the internal states of the ion. Such 
a detection is within the reach of the present ion trap technique$^{[26]}$, 
which presents us information about in which internal state the ion is
and causes negligible influence on the original process. Although they
are identical, the ions are distinguishable as long as the spacing
between any two of them is in the order of magnitude of $\mu m$$^{[11]}$, which
is much larger than the size of the trap ground state($10^{-9} m$)$^{[7]}$.
To carry out operation $\hat{U}$, we only need to implement Eq.(1) with
suitable choice of time and certain ion-pairs. However, to achieve operations
$\hat{M}^{(q)}_{i}$, the situation would be somewhat complicated. We take 
$\hat{M}^{(2)}_{1}$ as an example. If the first and second ions in the control pair 
(i.e., the ion pair acted as the control qubit) are in $|g>$ and $|e>$ 
respectively, the operation $\hat{M}^{(2)}_{1}$ will be the implementation of 
$\hat{U}(\frac {3\pi}{2})$($\hat{U}(\frac {\pi}{2})$) on the target ion-pair 
when the target ion-pair is in $|eg>(|ge>)$.

In summary, an approach with pairs of trapped ions to achieve the
Grover search has been proposed. As the logic states $|0>$ and $|1>$ are
degenerated in energy, the relative phase caused by the free evolution of
the states can be resorted to a global one, and thereby the detrimental
impact on the quantum interference in the quantum algorithm can be completely
diminished. We indicated that our scheme still meets
the requirement for the full Grover search although the iteration of the search is not
strictly periodic as the original Grover's approach. However, it is unclear 
in our scheme how to implement $\hat{M}^{(q)}_{i}$ for $q\geq 3$ more simply 
and efficiently. Moreover, the number of the ions 
required for quantum computing in our scheme is doubled compare to the former
approaches, which is in some sense uneconomic, particularly for the fact that 
it is an uneasy task for cooling down a few ions in the existing ion trap 
experiments. Nevertheless, our scheme is applicable and useful due to 
the advantages listed above. With the fast development of the ion 
trap technique,
we believe that much more entangled ions would be achieved in the linear ion 
trap in the near future.
Therefore, our scheme is a promising one for the hot-ion quantum computing.

\section{ Acknowledgement}

Valuable discussion with Xiwen Zhu and Kelin Gao is highly acknowledged.
The author is grateful to Andreas Buchleitner for his critical reading of this paper.  
The author also  sincerely thanks the referee for informing some new references.
The work is partly supported by the Chinese National Natural Science
Foundation.

\newpage
\begin{center}{\bf Captions of the figures}\end{center}

Fig.1~~ The probabilities of finding the marked state $|111>$ vs the number of 
iteration. As the amplitude of the state $|000>$ is the same as that of $|111>$ in the
iteration of the search,  we have
two readout results in this case. See text. The similar result can be obtained in
searching for other states.

Fig.2~~ The probabilities of finding the marked state $|1111>$ vs the number of 
iteration. The similar result can be obtained in searching for other states. 

Fig.3~~ The probabilities of finding the marked state $|11111>$ vs the number of 
iteration. The similar result can be obtained in searching for other states.

\newpage
\setlength{\unitlength}{0.240900pt}
\ifx\plotpoint\undefined\newsavebox{\plotpoint}\fi
\sbox{\plotpoint}{\rule[-0.200pt]{0.400pt}{0.400pt}}%
\begin{picture}(1500,900)(0,0)
\font\gnuplot=cmr10 at 10pt
\gnuplot
\sbox{\plotpoint}{\rule[-0.200pt]{0.400pt}{0.400pt}}%
\put(161.0,123.0){\rule[-0.200pt]{4.818pt}{0.400pt}}
\put(141,123){\makebox(0,0)[r]{0}}
\put(1419.0,123.0){\rule[-0.200pt]{4.818pt}{0.400pt}}
\put(161.0,216.0){\rule[-0.200pt]{4.818pt}{0.400pt}}
\put(141,216){\makebox(0,0)[r]{0.1}}
\put(1419.0,216.0){\rule[-0.200pt]{4.818pt}{0.400pt}}
\put(161.0,310.0){\rule[-0.200pt]{4.818pt}{0.400pt}}
\put(141,310){\makebox(0,0)[r]{0.2}}
\put(1419.0,310.0){\rule[-0.200pt]{4.818pt}{0.400pt}}
\put(161.0,403.0){\rule[-0.200pt]{4.818pt}{0.400pt}}
\put(141,403){\makebox(0,0)[r]{0.3}}
\put(1419.0,403.0){\rule[-0.200pt]{4.818pt}{0.400pt}}
\put(161.0,497.0){\rule[-0.200pt]{4.818pt}{0.400pt}}
\put(141,497){\makebox(0,0)[r]{0.4}}
\put(1419.0,497.0){\rule[-0.200pt]{4.818pt}{0.400pt}}
\put(161.0,590.0){\rule[-0.200pt]{4.818pt}{0.400pt}}
\put(141,590){\makebox(0,0)[r]{0.5}}
\put(1419.0,590.0){\rule[-0.200pt]{4.818pt}{0.400pt}}
\put(161.0,684.0){\rule[-0.200pt]{4.818pt}{0.400pt}}
\put(141,684){\makebox(0,0)[r]{0.6}}
\put(1419.0,684.0){\rule[-0.200pt]{4.818pt}{0.400pt}}
\put(161.0,777.0){\rule[-0.200pt]{4.818pt}{0.400pt}}
\put(141,777){\makebox(0,0)[r]{0.7}}
\put(1419.0,777.0){\rule[-0.200pt]{4.818pt}{0.400pt}}
\put(161.0,123.0){\rule[-0.200pt]{0.400pt}{4.818pt}}
\put(161,82){\makebox(0,0){0}}
\put(161.0,757.0){\rule[-0.200pt]{0.400pt}{4.818pt}}
\put(303.0,123.0){\rule[-0.200pt]{0.400pt}{4.818pt}}
\put(303,82){\makebox(0,0){2}}
\put(303.0,757.0){\rule[-0.200pt]{0.400pt}{4.818pt}}
\put(445.0,123.0){\rule[-0.200pt]{0.400pt}{4.818pt}}
\put(445,82){\makebox(0,0){4}}
\put(445.0,757.0){\rule[-0.200pt]{0.400pt}{4.818pt}}
\put(587.0,123.0){\rule[-0.200pt]{0.400pt}{4.818pt}}
\put(587,82){\makebox(0,0){6}}
\put(587.0,757.0){\rule[-0.200pt]{0.400pt}{4.818pt}}
\put(729.0,123.0){\rule[-0.200pt]{0.400pt}{4.818pt}}
\put(729,82){\makebox(0,0){8}}
\put(729.0,757.0){\rule[-0.200pt]{0.400pt}{4.818pt}}
\put(871.0,123.0){\rule[-0.200pt]{0.400pt}{4.818pt}}
\put(871,82){\makebox(0,0){10}}
\put(871.0,757.0){\rule[-0.200pt]{0.400pt}{4.818pt}}
\put(1013.0,123.0){\rule[-0.200pt]{0.400pt}{4.818pt}}
\put(1013,82){\makebox(0,0){12}}
\put(1013.0,757.0){\rule[-0.200pt]{0.400pt}{4.818pt}}
\put(1155.0,123.0){\rule[-0.200pt]{0.400pt}{4.818pt}}
\put(1155,82){\makebox(0,0){14}}
\put(1155.0,757.0){\rule[-0.200pt]{0.400pt}{4.818pt}}
\put(1297.0,123.0){\rule[-0.200pt]{0.400pt}{4.818pt}}
\put(1297,82){\makebox(0,0){16}}
\put(1297.0,757.0){\rule[-0.200pt]{0.400pt}{4.818pt}}
\put(1439.0,123.0){\rule[-0.200pt]{0.400pt}{4.818pt}}
\put(1439,82){\makebox(0,0){18}}
\put(1439.0,757.0){\rule[-0.200pt]{0.400pt}{4.818pt}}
\put(161.0,123.0){\rule[-0.200pt]{307.870pt}{0.400pt}}
\put(1439.0,123.0){\rule[-0.200pt]{0.400pt}{157.549pt}}
\put(161.0,777.0){\rule[-0.200pt]{307.870pt}{0.400pt}}
\put(40,450){\makebox(0,0){P}}
\put(800,21){\makebox(0,0){n}}
\put(800,839){\makebox(0,0){{\bf Fig.1}}}
\put(161.0,123.0){\rule[-0.200pt]{0.400pt}{157.549pt}}
\put(161,453){\usebox{\plotpoint}}
\multiput(161.58,448.73)(0.499,-1.165){139}{\rule{0.120pt}{1.030pt}}
\multiput(160.17,450.86)(71.000,-162.863){2}{\rule{0.400pt}{0.515pt}}
\multiput(232.58,288.00)(0.499,2.920){139}{\rule{0.120pt}{2.427pt}}
\multiput(231.17,288.00)(71.000,407.963){2}{\rule{0.400pt}{1.213pt}}
\multiput(303.58,688.03)(0.499,-3.797){139}{\rule{0.120pt}{3.125pt}}
\multiput(302.17,694.51)(71.000,-530.513){2}{\rule{0.400pt}{1.563pt}}
\multiput(374.58,164.00)(0.499,0.733){139}{\rule{0.120pt}{0.686pt}}
\multiput(373.17,164.00)(71.000,102.576){2}{\rule{0.400pt}{0.343pt}}
\multiput(445.58,268.00)(0.499,2.403){139}{\rule{0.120pt}{2.015pt}}
\multiput(444.17,268.00)(71.000,335.817){2}{\rule{0.400pt}{1.008pt}}
\multiput(516.58,601.20)(0.499,-1.929){139}{\rule{0.120pt}{1.638pt}}
\multiput(515.17,604.60)(71.000,-269.600){2}{\rule{0.400pt}{0.819pt}}
\multiput(587.58,335.00)(0.499,1.986){139}{\rule{0.120pt}{1.683pt}}
\multiput(586.17,335.00)(71.000,277.507){2}{\rule{0.400pt}{0.842pt}}
\multiput(658.58,607.17)(0.499,-2.545){139}{\rule{0.120pt}{2.128pt}}
\multiput(657.17,611.58)(71.000,-355.583){2}{\rule{0.400pt}{1.064pt}}
\multiput(729.58,253.22)(0.499,-0.712){139}{\rule{0.120pt}{0.669pt}}
\multiput(728.17,254.61)(71.000,-99.611){2}{\rule{0.400pt}{0.335pt}}
\multiput(800.58,155.00)(0.499,3.840){139}{\rule{0.120pt}{3.159pt}}
\multiput(799.17,155.00)(71.000,536.443){2}{\rule{0.400pt}{1.580pt}}
\multiput(871.58,688.11)(0.499,-2.863){139}{\rule{0.120pt}{2.382pt}}
\multiput(870.17,693.06)(71.000,-400.057){2}{\rule{0.400pt}{1.191pt}}
\multiput(942.58,293.00)(0.499,1.207){139}{\rule{0.120pt}{1.063pt}}
\multiput(941.17,293.00)(71.000,168.793){2}{\rule{0.400pt}{0.532pt}}
\multiput(1013.00,462.92)(1.631,-0.496){41}{\rule{1.391pt}{0.120pt}}
\multiput(1013.00,463.17)(68.113,-22.000){2}{\rule{0.695pt}{0.400pt}}
\multiput(1084.58,437.87)(0.499,-1.122){139}{\rule{0.120pt}{0.996pt}}
\multiput(1083.17,439.93)(71.000,-156.933){2}{\rule{0.400pt}{0.498pt}}
\multiput(1155.58,283.00)(0.499,2.976){139}{\rule{0.120pt}{2.472pt}}
\multiput(1154.17,283.00)(71.000,415.870){2}{\rule{0.400pt}{1.236pt}}
\multiput(1226.58,691.19)(0.499,-3.748){139}{\rule{0.120pt}{3.086pt}}
\multiput(1225.17,697.60)(71.000,-523.595){2}{\rule{0.400pt}{1.543pt}}
\multiput(1297.58,174.00)(0.499,0.910){139}{\rule{0.120pt}{0.827pt}}
\multiput(1296.17,174.00)(71.000,127.284){2}{\rule{0.400pt}{0.413pt}}
\multiput(1368.58,303.00)(0.499,1.851){139}{\rule{0.120pt}{1.576pt}}
\multiput(1367.17,303.00)(71.000,258.729){2}{\rule{0.400pt}{0.788pt}}
\end{picture}

\setlength{\unitlength}{0.240900pt}
\ifx\plotpoint\undefined\newsavebox{\plotpoint}\fi
\begin{picture}(1500,900)(0,40)
\font\gnuplot=cmr10 at 10pt
\gnuplot
\sbox{\plotpoint}{\rule[-0.200pt]{0.400pt}{0.400pt}}%
\put(161.0,123.0){\rule[-0.200pt]{4.818pt}{0.400pt}}
\put(141,123){\makebox(0,0)[r]{0.1}}
\put(1419.0,123.0){\rule[-0.200pt]{4.818pt}{0.400pt}}
\put(161.0,216.0){\rule[-0.200pt]{4.818pt}{0.400pt}}
\put(141,216){\makebox(0,0)[r]{0.2}}
\put(1419.0,216.0){\rule[-0.200pt]{4.818pt}{0.400pt}}
\put(161.0,310.0){\rule[-0.200pt]{4.818pt}{0.400pt}}
\put(141,310){\makebox(0,0)[r]{0.3}}
\put(1419.0,310.0){\rule[-0.200pt]{4.818pt}{0.400pt}}
\put(161.0,403.0){\rule[-0.200pt]{4.818pt}{0.400pt}}
\put(141,403){\makebox(0,0)[r]{0.4}}
\put(1419.0,403.0){\rule[-0.200pt]{4.818pt}{0.400pt}}
\put(161.0,497.0){\rule[-0.200pt]{4.818pt}{0.400pt}}
\put(141,497){\makebox(0,0)[r]{0.5}}
\put(1419.0,497.0){\rule[-0.200pt]{4.818pt}{0.400pt}}
\put(161.0,590.0){\rule[-0.200pt]{4.818pt}{0.400pt}}
\put(141,590){\makebox(0,0)[r]{0.6}}
\put(1419.0,590.0){\rule[-0.200pt]{4.818pt}{0.400pt}}
\put(161.0,684.0){\rule[-0.200pt]{4.818pt}{0.400pt}}
\put(141,684){\makebox(0,0)[r]{0.7}}
\put(1419.0,684.0){\rule[-0.200pt]{4.818pt}{0.400pt}}
\put(161.0,777.0){\rule[-0.200pt]{4.818pt}{0.400pt}}
\put(141,777){\makebox(0,0)[r]{0.8}}
\put(1419.0,777.0){\rule[-0.200pt]{4.818pt}{0.400pt}}
\put(161.0,123.0){\rule[-0.200pt]{0.400pt}{4.818pt}}
\put(161,82){\makebox(0,0){0}}
\put(161.0,757.0){\rule[-0.200pt]{0.400pt}{4.818pt}}
\put(303.0,123.0){\rule[-0.200pt]{0.400pt}{4.818pt}}
\put(303,82){\makebox(0,0){2}}
\put(303.0,757.0){\rule[-0.200pt]{0.400pt}{4.818pt}}
\put(445.0,123.0){\rule[-0.200pt]{0.400pt}{4.818pt}}
\put(445,82){\makebox(0,0){4}}
\put(445.0,757.0){\rule[-0.200pt]{0.400pt}{4.818pt}}
\put(587.0,123.0){\rule[-0.200pt]{0.400pt}{4.818pt}}
\put(587,82){\makebox(0,0){6}}
\put(587.0,757.0){\rule[-0.200pt]{0.400pt}{4.818pt}}
\put(729.0,123.0){\rule[-0.200pt]{0.400pt}{4.818pt}}
\put(729,82){\makebox(0,0){8}}
\put(729.0,757.0){\rule[-0.200pt]{0.400pt}{4.818pt}}
\put(871.0,123.0){\rule[-0.200pt]{0.400pt}{4.818pt}}
\put(871,82){\makebox(0,0){10}}
\put(871.0,757.0){\rule[-0.200pt]{0.400pt}{4.818pt}}
\put(1013.0,123.0){\rule[-0.200pt]{0.400pt}{4.818pt}}
\put(1013,82){\makebox(0,0){12}}
\put(1013.0,757.0){\rule[-0.200pt]{0.400pt}{4.818pt}}
\put(1155.0,123.0){\rule[-0.200pt]{0.400pt}{4.818pt}}
\put(1155,82){\makebox(0,0){14}}
\put(1155.0,757.0){\rule[-0.200pt]{0.400pt}{4.818pt}}
\put(1297.0,123.0){\rule[-0.200pt]{0.400pt}{4.818pt}}
\put(1297,82){\makebox(0,0){16}}
\put(1297.0,757.0){\rule[-0.200pt]{0.400pt}{4.818pt}}
\put(1439.0,123.0){\rule[-0.200pt]{0.400pt}{4.818pt}}
\put(1439,82){\makebox(0,0){18}}
\put(1439.0,757.0){\rule[-0.200pt]{0.400pt}{4.818pt}}
\put(161.0,123.0){\rule[-0.200pt]{307.870pt}{0.400pt}}
\put(1439.0,123.0){\rule[-0.200pt]{0.400pt}{157.549pt}}
\put(161.0,777.0){\rule[-0.200pt]{307.870pt}{0.400pt}}
\put(40,450){\makebox(0,0){P}}
\put(800,21){\makebox(0,0){n}}
\put(800,839){\makebox(0,0){{\bf Fig.2}}}
\put(161.0,123.0){\rule[-0.200pt]{0.400pt}{157.549pt}}
\put(161,263){\usebox{\plotpoint}}
\multiput(161.00,263.58)(0.602,0.499){115}{\rule{0.581pt}{0.120pt}}
\multiput(161.00,262.17)(69.793,59.000){2}{\rule{0.291pt}{0.400pt}}
\multiput(232.58,322.00)(0.499,1.547){139}{\rule{0.120pt}{1.334pt}}
\multiput(231.17,322.00)(71.000,216.232){2}{\rule{0.400pt}{0.667pt}}
\multiput(303.58,541.00)(0.499,1.335){139}{\rule{0.120pt}{1.165pt}}
\multiput(302.17,541.00)(71.000,186.582){2}{\rule{0.400pt}{0.582pt}}
\multiput(374.58,726.36)(0.499,-0.973){281}{\rule{0.120pt}{0.877pt}}
\multiput(373.17,728.18)(142.000,-274.179){2}{\rule{0.400pt}{0.439pt}}
\multiput(516.58,454.00)(0.499,1.950){139}{\rule{0.120pt}{1.655pt}}
\multiput(515.17,454.00)(71.000,272.565){2}{\rule{0.400pt}{0.827pt}}
\multiput(587.58,716.25)(0.499,-4.031){139}{\rule{0.120pt}{3.311pt}}
\multiput(586.17,723.13)(71.000,-563.127){2}{\rule{0.400pt}{1.656pt}}
\multiput(658.58,160.00)(0.499,1.009){139}{\rule{0.120pt}{0.906pt}}
\multiput(657.17,160.00)(71.000,141.120){2}{\rule{0.400pt}{0.453pt}}
\multiput(729.58,299.05)(0.499,-1.066){139}{\rule{0.120pt}{0.951pt}}
\multiput(728.17,301.03)(71.000,-149.027){2}{\rule{0.400pt}{0.475pt}}
\multiput(800.00,150.92)(1.559,-0.496){43}{\rule{1.335pt}{0.120pt}}
\multiput(800.00,151.17)(68.230,-23.000){2}{\rule{0.667pt}{0.400pt}}
\multiput(871.58,129.00)(0.499,0.853){139}{\rule{0.120pt}{0.782pt}}
\multiput(870.17,129.00)(71.000,119.378){2}{\rule{0.400pt}{0.391pt}}
\multiput(942.58,250.00)(0.499,1.023){139}{\rule{0.120pt}{0.917pt}}
\multiput(941.17,250.00)(71.000,143.097){2}{\rule{0.400pt}{0.458pt}}
\multiput(1013.58,395.00)(0.499,0.853){139}{\rule{0.120pt}{0.782pt}}
\multiput(1012.17,395.00)(71.000,119.378){2}{\rule{0.400pt}{0.391pt}}
\multiput(1084.58,513.64)(0.499,-0.584){139}{\rule{0.120pt}{0.568pt}}
\multiput(1083.17,514.82)(71.000,-81.822){2}{\rule{0.400pt}{0.284pt}}
\multiput(1155.58,433.00)(0.499,1.533){139}{\rule{0.120pt}{1.323pt}}
\multiput(1154.17,433.00)(71.000,214.255){2}{\rule{0.400pt}{0.661pt}}
\multiput(1226.58,643.88)(0.499,-1.724){139}{\rule{0.120pt}{1.475pt}}
\multiput(1225.17,646.94)(71.000,-240.939){2}{\rule{0.400pt}{0.737pt}}
\multiput(1297.58,400.21)(0.499,-1.625){139}{\rule{0.120pt}{1.396pt}}
\multiput(1296.17,403.10)(71.000,-227.103){2}{\rule{0.400pt}{0.698pt}}
\end{picture}

\setlength{\unitlength}{0.240900pt}
\ifx\plotpoint\undefined\newsavebox{\plotpoint}\fi
\begin{picture}(1500,900)(0,80)
\font\gnuplot=cmr10 at 10pt
\gnuplot
\sbox{\plotpoint}{\rule[-0.200pt]{0.400pt}{0.400pt}}%
\put(161.0,123.0){\rule[-0.200pt]{4.818pt}{0.400pt}}
\put(141,123){\makebox(0,0)[r]{0}}
\put(1419.0,123.0){\rule[-0.200pt]{4.818pt}{0.400pt}}
\put(161.0,216.0){\rule[-0.200pt]{4.818pt}{0.400pt}}
\put(141,216){\makebox(0,0)[r]{0.1}}
\put(1419.0,216.0){\rule[-0.200pt]{4.818pt}{0.400pt}}
\put(161.0,310.0){\rule[-0.200pt]{4.818pt}{0.400pt}}
\put(141,310){\makebox(0,0)[r]{0.2}}
\put(1419.0,310.0){\rule[-0.200pt]{4.818pt}{0.400pt}}
\put(161.0,403.0){\rule[-0.200pt]{4.818pt}{0.400pt}}
\put(141,403){\makebox(0,0)[r]{0.3}}
\put(1419.0,403.0){\rule[-0.200pt]{4.818pt}{0.400pt}}
\put(161.0,497.0){\rule[-0.200pt]{4.818pt}{0.400pt}}
\put(141,497){\makebox(0,0)[r]{0.4}}
\put(1419.0,497.0){\rule[-0.200pt]{4.818pt}{0.400pt}}
\put(161.0,590.0){\rule[-0.200pt]{4.818pt}{0.400pt}}
\put(141,590){\makebox(0,0)[r]{0.5}}
\put(1419.0,590.0){\rule[-0.200pt]{4.818pt}{0.400pt}}
\put(161.0,684.0){\rule[-0.200pt]{4.818pt}{0.400pt}}
\put(141,684){\makebox(0,0)[r]{0.6}}
\put(1419.0,684.0){\rule[-0.200pt]{4.818pt}{0.400pt}}
\put(161.0,777.0){\rule[-0.200pt]{4.818pt}{0.400pt}}
\put(141,777){\makebox(0,0)[r]{0.7}}
\put(1419.0,777.0){\rule[-0.200pt]{4.818pt}{0.400pt}}
\put(161.0,123.0){\rule[-0.200pt]{0.400pt}{4.818pt}}
\put(161,82){\makebox(0,0){0}}
\put(161.0,757.0){\rule[-0.200pt]{0.400pt}{4.818pt}}
\put(303.0,123.0){\rule[-0.200pt]{0.400pt}{4.818pt}}
\put(303,82){\makebox(0,0){2}}
\put(303.0,757.0){\rule[-0.200pt]{0.400pt}{4.818pt}}
\put(445.0,123.0){\rule[-0.200pt]{0.400pt}{4.818pt}}
\put(445,82){\makebox(0,0){4}}
\put(445.0,757.0){\rule[-0.200pt]{0.400pt}{4.818pt}}
\put(587.0,123.0){\rule[-0.200pt]{0.400pt}{4.818pt}}
\put(587,82){\makebox(0,0){6}}
\put(587.0,757.0){\rule[-0.200pt]{0.400pt}{4.818pt}}
\put(729.0,123.0){\rule[-0.200pt]{0.400pt}{4.818pt}}
\put(729,82){\makebox(0,0){8}}
\put(729.0,757.0){\rule[-0.200pt]{0.400pt}{4.818pt}}
\put(871.0,123.0){\rule[-0.200pt]{0.400pt}{4.818pt}}
\put(871,82){\makebox(0,0){10}}
\put(871.0,757.0){\rule[-0.200pt]{0.400pt}{4.818pt}}
\put(1013.0,123.0){\rule[-0.200pt]{0.400pt}{4.818pt}}
\put(1013,82){\makebox(0,0){12}}
\put(1013.0,757.0){\rule[-0.200pt]{0.400pt}{4.818pt}}
\put(1155.0,123.0){\rule[-0.200pt]{0.400pt}{4.818pt}}
\put(1155,82){\makebox(0,0){14}}
\put(1155.0,757.0){\rule[-0.200pt]{0.400pt}{4.818pt}}
\put(1297.0,123.0){\rule[-0.200pt]{0.400pt}{4.818pt}}
\put(1297,82){\makebox(0,0){16}}
\put(1297.0,757.0){\rule[-0.200pt]{0.400pt}{4.818pt}}
\put(1439.0,123.0){\rule[-0.200pt]{0.400pt}{4.818pt}}
\put(1439,82){\makebox(0,0){18}}
\put(1439.0,757.0){\rule[-0.200pt]{0.400pt}{4.818pt}}
\put(161.0,123.0){\rule[-0.200pt]{307.870pt}{0.400pt}}
\put(1439.0,123.0){\rule[-0.200pt]{0.400pt}{157.549pt}}
\put(161.0,777.0){\rule[-0.200pt]{307.870pt}{0.400pt}}
\put(40,450){\makebox(0,0){P}}
\put(800,21){\makebox(0,0){n}}
\put(800,839){\makebox(0,0){{\bf Fig.3}}}
\put(161.0,123.0){\rule[-0.200pt]{0.400pt}{157.549pt}}
\put(161,288){\usebox{\plotpoint}}
\multiput(161.00,286.93)(5.374,-0.485){11}{\rule{4.157pt}{0.117pt}}
\multiput(161.00,287.17)(62.372,-7.000){2}{\rule{2.079pt}{0.400pt}}
\multiput(232.58,281.00)(0.499,1.759){139}{\rule{0.120pt}{1.503pt}}
\multiput(231.17,281.00)(71.000,245.881){2}{\rule{0.400pt}{0.751pt}}
\multiput(303.58,521.35)(0.499,-2.488){139}{\rule{0.120pt}{2.083pt}}
\multiput(302.17,525.68)(71.000,-347.676){2}{\rule{0.400pt}{1.042pt}}
\multiput(374.58,178.00)(0.499,3.826){139}{\rule{0.120pt}{3.148pt}}
\multiput(373.17,178.00)(71.000,534.466){2}{\rule{0.400pt}{1.574pt}}
\multiput(445.58,710.47)(0.499,-2.453){139}{\rule{0.120pt}{2.055pt}}
\multiput(444.17,714.73)(71.000,-342.735){2}{\rule{0.400pt}{1.027pt}}
\multiput(516.58,372.00)(0.499,0.839){139}{\rule{0.120pt}{0.770pt}}
\multiput(515.17,372.00)(71.000,117.401){2}{\rule{0.400pt}{0.385pt}}
\multiput(587.58,486.54)(0.499,-1.221){139}{\rule{0.120pt}{1.075pt}}
\multiput(586.17,488.77)(71.000,-170.770){2}{\rule{0.400pt}{0.537pt}}
\multiput(658.58,318.00)(0.499,0.754){139}{\rule{0.120pt}{0.703pt}}
\multiput(657.17,318.00)(71.000,105.541){2}{\rule{0.400pt}{0.351pt}}
\multiput(729.00,423.92)(3.326,-0.492){19}{\rule{2.682pt}{0.118pt}}
\multiput(729.00,424.17)(65.434,-11.000){2}{\rule{1.341pt}{0.400pt}}
\multiput(800.58,414.00)(0.499,1.731){139}{\rule{0.120pt}{1.480pt}}
\multiput(799.17,414.00)(71.000,241.928){2}{\rule{0.400pt}{0.740pt}}
\multiput(871.58,654.28)(0.499,-1.299){139}{\rule{0.120pt}{1.137pt}}
\multiput(870.17,656.64)(71.000,-181.641){2}{\rule{0.400pt}{0.568pt}}
\multiput(942.58,469.53)(0.499,-1.526){139}{\rule{0.120pt}{1.317pt}}
\multiput(941.17,472.27)(71.000,-213.267){2}{\rule{0.400pt}{0.658pt}}
\multiput(1013.58,259.00)(0.499,0.790){139}{\rule{0.120pt}{0.731pt}}
\multiput(1012.17,259.00)(71.000,110.483){2}{\rule{0.400pt}{0.365pt}}
\multiput(1084.58,371.00)(0.499,1.023){139}{\rule{0.120pt}{0.917pt}}
\multiput(1083.17,371.00)(71.000,143.097){2}{\rule{0.400pt}{0.458pt}}
\multiput(1155.58,516.00)(0.499,1.193){139}{\rule{0.120pt}{1.052pt}}
\multiput(1154.17,516.00)(71.000,166.816){2}{\rule{0.400pt}{0.526pt}}
\multiput(1226.58,676.75)(0.499,-2.368){139}{\rule{0.120pt}{1.987pt}}
\multiput(1225.17,680.88)(71.000,-330.875){2}{\rule{0.400pt}{0.994pt}}
\multiput(1297.58,347.01)(0.499,-0.775){139}{\rule{0.120pt}{0.720pt}}
\multiput(1296.17,348.51)(71.000,-108.506){2}{\rule{0.400pt}{0.360pt}}
\end{picture}

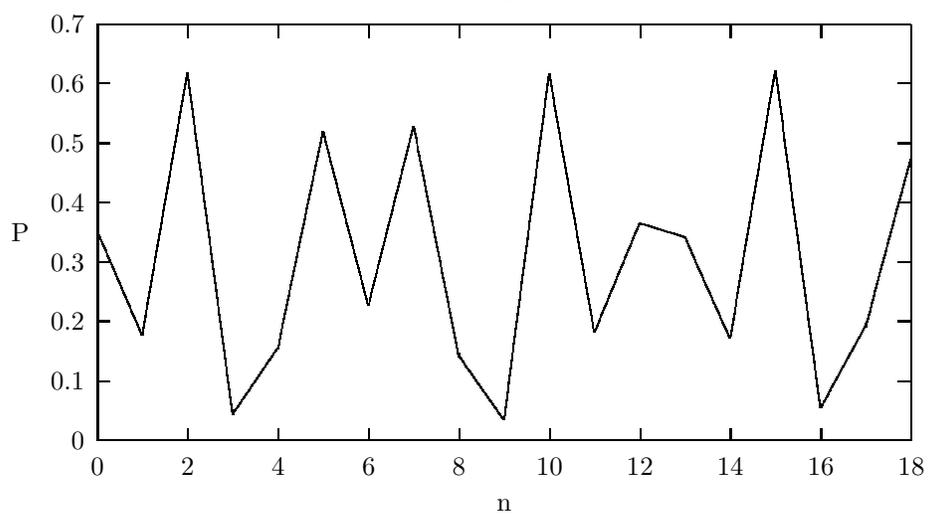
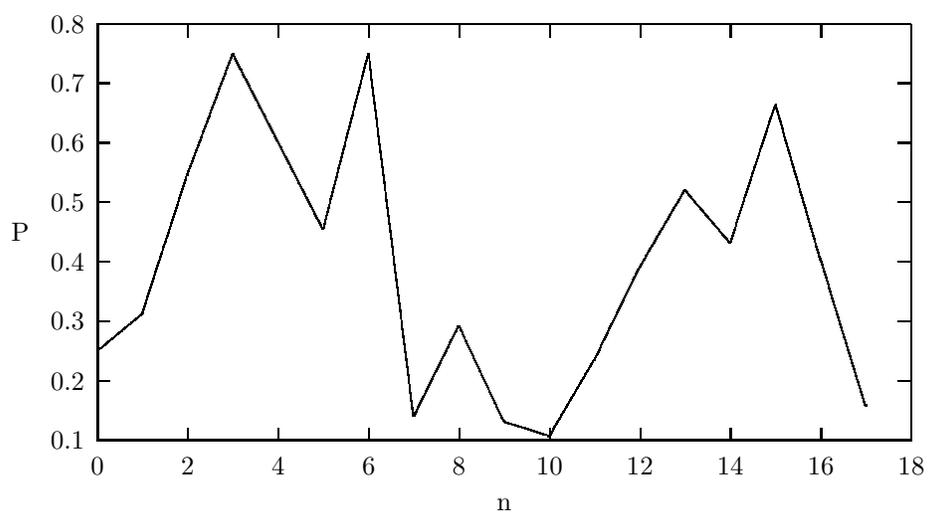
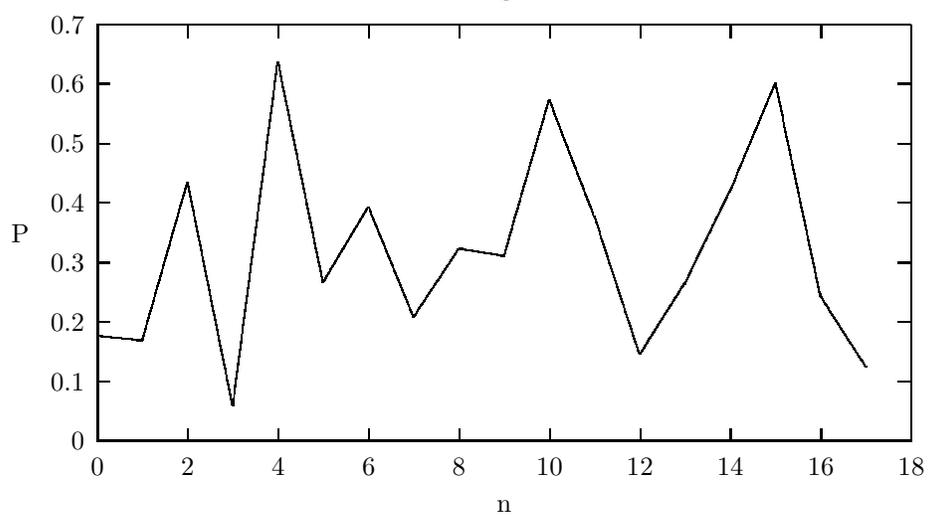
\end{document}